\documentclass[arguments]{aastex631}

\usepackage{soul}

\begin{document}

\title{A New Component from the Quiet Sun: Synchrotron Radiation from Galactic Cosmic-Ray Electrons}

\author{Elena Orlando}
\affiliation{University of Trieste and National Institute for Nuclear Physics (INFN), Italy}
\affiliation{Kavli Institute for Particle Astrophysics and Cosmology and Hansen Experimental Physics Laboratory, Stanford University, USA}

\author{Vahe' Petrosian}
\affiliation{Kavli Institute for Particle Astrophysics and Cosmology and Hansen Experimental Physics Laboratory, Stanford University, USA}

\author{Andrew Strong}
\affiliation{The Max Planck Institute for Extraterrestrial Physics, Germany}



\begin{abstract}
The quiet Sun, i.e. in its non-flaring state or non-flaring regions, emits thermal radiation from radio to ultraviolet. The quiet Sun produces also non-thermal radiation observed in gamma rays due to interactions of Galactic Cosmic Rays (GCR) with the solar gas and photons. \\
We report on a new component: the synchrotron emission by GCR electrons in the solar magnetic field. To the best of our knowledge this is the first time this emission has been theoretically claimed and modeled. 
We find that the measured GCR electrons with energies from tens of GeV to a few TeV produce synchrotron emission in X-rays, which is a few orders of magnitude lower than current upper limits of the quiet Sun set by RHESSI and FOXSI. For a radially decreasing solar magnetic field we find the expected synchrotron intensity to be almost constant in the solar disk, to peak in the close proximity of the Sun, and to quickly drop away from the Sun. We also estimate the synchrotron emission from radio to gamma rays and we compare it with current observations, especially with LOFAR. While it is negligible from radio to UV compared to the solar thermal radiation, this emission can potentially be observed at high energies with NuSTAR and more promising future FOXSI observations. 
This could potentially allow for constraining CR densities and magnetic-field intensities at the Sun. This study provides a more complete description and a possible new way for understanding the quite Sun and its environment.
\end{abstract}



\section{Introduction} \label{sec:intro}
\label{intro}
The quiet Sun, defined as its non-flaring state or non-flaring regions, is known to emit radiation from radio to gamma rays. While most of the radiation from the quiet Sun has thermal origin, at high energies it is produced by accelerated particles.
In fact, the quiet gamma-ray emission above a few tens of MeV originates from interactions of Galactic Cosmic Rays (GCR) with the solar gas and the solar photons. In more detail, a disk component is supposed to be due to pion decay of GCR hadrons interacting with the solar gas \citep{Seckel91, Thompson, Mazziotta, Li, Becker, Guti, Hudson20, Nibl, Zhou}, while a spatially extended component is supposed to be due to inverse-Compton scattering of GCR electrons on the solar photon field \citep{Orlando2006, Moskalenko, Stellarics, Lai}. Both components have been observed with the Energetic Gamma Ray Experiment Telescope (EGRET) first \citep{Orlando2008}, and with Fermi Large Area telescope later \citep{Abdo2011, Barbiellini, Bartoli, Linden, Linden2020, Ng, Tang}.
In X-rays above a few keV (i.e. hard X-rays), the solar corona on its non-flaring state or non-flaring regions emits at millions of degrees, while the underlying
chromosphere and photosphere is much cooler.
This is know as the coronal heating problem, which is still one of the fundamental unanswered questions in
solar physics (see e.g. the review by
\cite{Klimchuk} and references therein). Nanoflares are among current possible explanations of the corona heating problem \citep{Parker}.
The Reuven Ramaty High Energy Solar Spectroscopic Imager (RHESSI) \citep{RHESSImission}, launched in 2003 and decommissioned in 2018, imaged the Sun and solar flares and provided spatially resolved spectroscopy with high spectral resolution from $\sim$3~keV to several MeV. 
RHESSI, the Focusing Optics X-ray Solar Imager (FOXSI) \citep{FOXSImission}, and the Nuclear Spectroscopic Telescope Array (NuSTAR) \citep{NUSTARmission} have recently provided solar observations above a few keV, with RHESSI  posing stringent upper limits to the quiet solar emission between 3 and 200 keV \citep{Hannah2008, RHESSI2010} using almost 12 days of quiescent solar observations during solar minimum. These are still the deepest limits for solar hard X-ray emission yet reported for this energy range. At 3-6 keV and at 6-12 keV these observations set upper limits at 3.4 $\times$ 10$^{-2}$ photons s$^{-1}$ cm$^{-2}$ keV$^{-1}$  and 9.6 $\times$ 10$^{-4}$ photons s$^{-1}$ cm$^{-2}$ keV$^{-1}$, which have recently been confirmed by FOXSI \citep{Foxsi} with just a few minutes of observations. Moreover, the FOXSI rocket mission assessed the hard X-ray flux of a quiescent solar region during a substantially high solar activity period for the first time. In the 5-10 keV energy range FOXSI-2 reached an upper limit of 4.5 $\times$ 10$^{-2}$ photons s$^{-1}$ cm$^{-2}$ keV$^{-1}$ during a period of high solar activity, and FOXSI-3 reached an upper limit of 9.3 - 6~$\times$10$^{-4}$ photons s$^{-1}$ cm$^{-2}$ keV$^{-1}$ during a period of low solar activity. 
In the radio band the quiet solar emission is interpreted to be produced by thermal Bremsstrahlung  emission  \citep[e.g.][]{Selhorst}. 
In the optical band, in the UV band, and in the soft X-ray band (i.e. below a few keV) the solar emission is mostly of thermal origin.
In the X-ray energy range, even during periods of low solar activity, the Sun's atmosphere is still filled with small scale events like jets, flares, and minifilament eruptions.
Here we introduce a new emission component of the quiet Sun from gamma rays to radio, with particular emphasis in the X-ray band thanks to RHESSI and FOXSI upper limits. This  non-thermal emission component is produced by synchrotron radiation of GCR electrons in the solar magnetic field. Section 2 explains our modeling and compares our estimates at high energies with the current data of the quiet Sun in X-rays by RHESSI and FOXSI. Section 3 discusses our predictions for future observations in X-rays and it reports on our estimates of this synchrotron component from radio to UV in light of current and future observations.

\section{X-ray Synchrotron modeling and observations}
In this section we report our calculations of the synchrotron emission by GCR in the solar magnetic field and we compare it with current observations.\\
The synchrotron emissivity 
is calculated numerically on a frequency grid and spatial grid in the Heliosphere (x, y, $\nu$) centred on the Sun. The spectrum and distribution of the synchrotron emissivity depend on the intensity of the magnetic field and the spectrum and distribution of CR electrons at each point in the grid. 
For a randomly oriented magnetic field with intensity $B$ and for an electron with Lorentz factor $\gamma$, the emissivity is isotropic and obtained with the formulation given by \cite{Ghisellini}:\\
\begin{equation}
\epsilon(\nu, \gamma)= C\ x^2 [K_{4/3}K_{1/3} -{3\over5} x (K_{4/3}K_{4/3} -K_{1/3}K_{1/3})]
\label{eq1}
\end{equation}
with $x=\nu/\nu_c$,  $\nu_c={3\over2\pi} {e\over mc} B \gamma^2$,
$C={2\sqrt3} {e^3\over mc^2} B$ erg s$^{-1}$ Hz$^{-1}$,
and $ K_{4/3}$, $K_{1/3}$ Bessel functions, computed using the GNU Scientific Library\footnote{http://www.gnu.org/software/gsl}.
From there, the relation between the characteristic synchrotron photon energy E, the CR electron energy E$_e$, and the intensity of the magnetic field B, follows the analytical approximation:
\begin{equation}
E~\approx ~67~B~(E_e~/~10^{12})^2
\label{eq2}
\end{equation}
with E in keV, E$_e$ in eV, and B in G.
To obtain the synchrotron intensity we integrate  the calculated emissivity over the line-of-sight.
Hence, following \cite{SO2011} and \cite{OS2013} the synchrotron intensity along a line-of sight $s$, at frequency $\nu$, for a given isotropic distribution and spectrum of CR electrons, n$_{\gamma}$, is given by: 
\begin{equation}
I(\nu)=\int \int \epsilon(\nu, \gamma)~n_{\gamma} ~d\gamma ~ds
\label{eq3}
\end{equation}
Because of the high-energy electrons involved for the synchrotron modeling in X-rays and gamma rays, which are produced by electrons above a few tens of GeV, we can consider the solar modulation of GCR to be negligible. 
\begin{figure} 
\centering
\includegraphics[width=0.6\textwidth, angle=0]{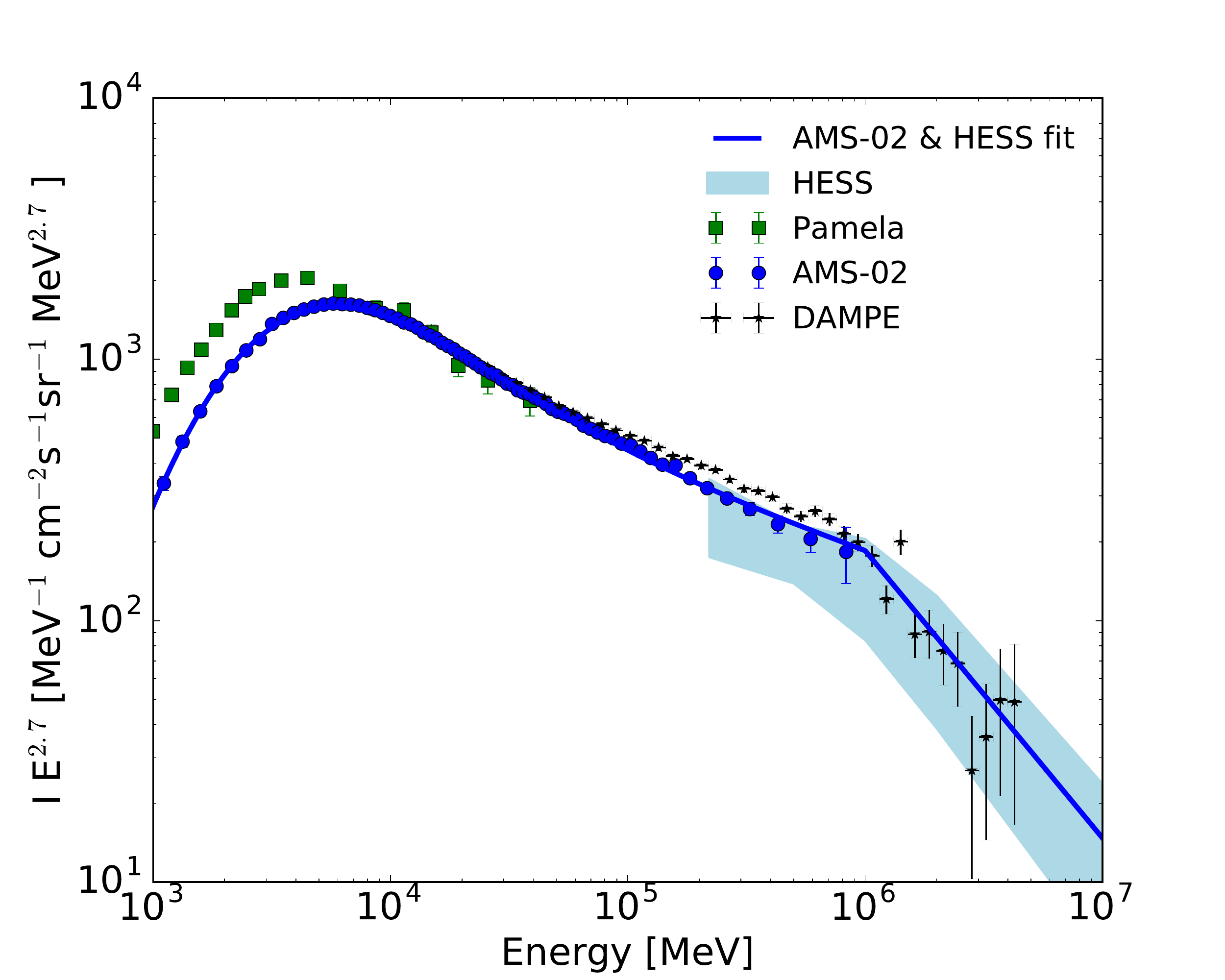}
\caption{Modelled electron spectrum (blue solid line) compared with data. 
The model is fitted to the AMS-02 \citep{AMS02} electron spectrum for the period of 2013 (blue points) and to HESS measurements (bluish region and black triangles). The bluish region identifies the HESS measurements uncertainty. The electron spectrum measured by PAMELA \citep{Pamela}  is also shown (green squares), with measurements by DAMPE (black stars) \citep{DAMPE}.
The plotted spectrum is used to calculate the synchrotron emission.
\label{figure1}}
\end{figure}
For our synchrotron modeling we assume the GCR electron spectrum (electrons plus positrons) as measured by the Alpha Magnetic Spectrometer (AMS-02) \citep{AMS02} and by the High Energy Stereoscopic System (HESS)\footnote{https://www.mpi-hd.mpg.de/hfm/HESS/pages/home/som/2017/09/}. 
The assumed GCR electron spectrum is shown in Figure \ref{figure1}. 
The blue solid line shows the GCR electron spectrum that fits AMS-02 and HESS data. The plot also illustrates the AMS-02 \citep{AMS02}, the Antimatter Matter Exploration and Light-nuclei Astrophysics (PAMELA) \citep{Pamela}, the Dark Matter Particle Explorer (DAMPE) \citep{DAMPE}, and the HESS measurements. 
Below several hundreds of MeV, electrons are affected by solar modulation, which is why PAMELA and AMS-02 measurements differ.  DAMPE measurements are somehow higher than AMS-02 in the region of interest for X-ray synchrotron emission, which makes our calculations conservative. However, here we neglect GCR energy losses, with the result that our estimates of the synchrotron intensity should be treated as an upper limit (see Section 3.3 for a deeper discussion). \\
For calculating the synchrotron emission we assume given magnetic field models as in \cite{POS} and described in the following.  
It is believed that the intensity of the magnetic field in the heliosphere follows a Parker spiral law with $B(r)$~$\propto$~r$^{-\delta}$ and $\delta \approx$~2 with $r$ distance from the Sun, as confirmed by recent observations by Parker Solar Probe \citep{Badman}. 
We consider two regions for the magnetic field: the inner Heliosphere from 0.1 $<$ r/AU $<$ 1, and the region closer to the Sun with r/AU $<$ 0.1. For the former we assume the intensity of the magnetic field that fits recent observations by the Parker Solar Probe extending to about 20~$R_{sun}$ (or r/AU $=$ 0.1). For the latter we assume two different models as in the following. A model based on \cite{GY11} (we called it GY11), which derived the intensity variation in the region 5 $<$ r/AU $<$ 25 by using observations of the corona mass ejections; a model based on \cite{Patzold} (we called it Patzold), which agrees with Parker Solar Probe observations and steepens at the photosphere. Due to the current limited knowledge of the magnetic field at the Sun, we consider the two models above to bracket the uncertainty. Also, the different intensity between the two models accounts for possible different solar conditions.
After assuming continuation in the point of transition between models, we obtain the following parametrization for the intensity of the magnetic field: 

\begin{equation}
  B(r)=\left\{
    \begin{array}{lll}
      1.0~(r/R_{sun})^{-1.9}, & \mbox{Parker Solar Probe for $0.1<r/AU<1.0$}.\\
      0.31~(r/R_{sun})^{-1.5}, & \mbox{GY11, for $r/AU<0.1$}.\\
      8.4~(r/R_{sun})^{-2.6}, & \mbox{Patzold, for $r/AU<0.1$}.\\
    \end{array}
  \right.
  \\
\end{equation}
with B(r) in Gauss. The intensity of the magnetic field as a function of the distance from the Sun in solar radii is reported in Figure~\ref{figure2}. 
\begin{figure} 
 \centering
\includegraphics[width=0.5\textwidth, angle=0]{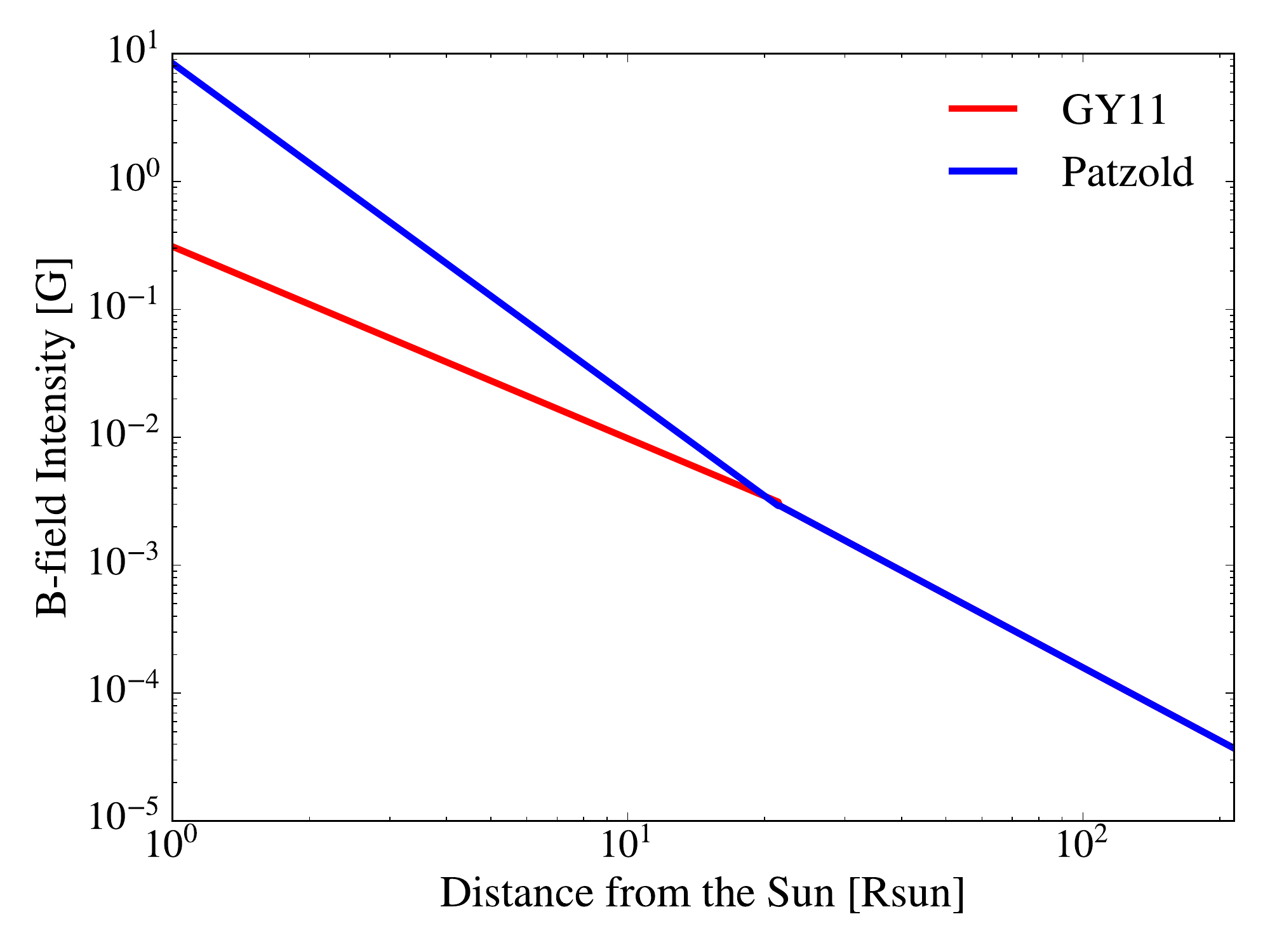}
\caption{Assumed magnetic field intensity as a function of the distance from the Sun in solar radii. Lines are analytical fits to the observed magnetic field as explained in the text and in \cite{POS}. The red line (GY11) shows the radial intensity based on \cite{GY11} and Parker Solar Probe measurements \citep{Badman}. The blue line (Patzold) is based on the magnetic field as in \cite{Patzold} and \cite{Badman}. For r/AU$>$0.1 the GY11 and Patzold models assume a magnetic field intensity based on the Parker Solar Probe measurements. 
\label{figure2}}
\end{figure}\\
Close to the Sun, in a magnetic field of a few G, synchrotron emission in the RHESSI energy band is produced mainly by electrons from a few tens of GeV to a few TeV (see e.g. eq.\ref{eq2}).
We calculate the synchrotron emission up to 100 MeV where\footnote{Synchrotron emission has a maximum energy photon around 200 MeV, above which radiation reaction becomes dominant and particles lose energy in a few orbits.}, due to the relatively low magnetic field strength, we can consider to be in the classical regime\footnote{Quantum mechanical effects become important when (3/2)~$\times$~$\gamma$~$\times$~(B/B$_{cr}$)~$>$~1,
with B$_{cr}$~=~4~$\times$~10$^{12}$~G (see \cite{Petrosian87}). For example, 
for B~=~10~G, $\gamma$ would need to be $>$~10$^{11}$.}.  Note that GCR electrons below several TeV are constrained by direct measurements. Because GCR electron direct measurements extends up to $\sim$10 TeV, which produce synchrotron emission up to a few tens of MeV for a magnetic field of several G, the electron spectrum is extrapolated up to 100 TeV.
Figure~\ref{figure3} shows the calculated synchrotron spectral flux for the two magnetic field models defined above: GY11 (black lines) and Patzold (blue lines). 
By comparing the two models we see that the synchrotron flux is very sensitive to the magnetic field intensity, which is not surprising because it scales as  B$^{(p+1)/2}$, where $p\sim 3$ is the spectral index of the electrons. Models are compared with RHESSI \citep{RHESSI2010} and FOXSI \citep{Foxsi} upper limits of the quiet Sun. The plotted observed upper limits were integrated for the entire solid angle of the Sun (private communications with the authors). 
We see that synchrotron emission produces X-rays in the energy band where upper limits to the quiet solar emission have been obtained. The expected synchrotron emission is a few orders of magnitude lower than current upper limits of the quiet Sun, which makes it promising to be observed in future. The same figure shows dashed lines that are obtained by integrating the emission in the entire solar disk, and solid lines that are obtained by integrating the emission over a circular region of 1 degree radius from the center of the Sun. Note that the calculated synchrotron flux including the region outside the solar disk is  larger than the calculated flux on the disk alone. 
\begin{figure} 
\centering
\includegraphics[width=0.6\textwidth, angle=0]{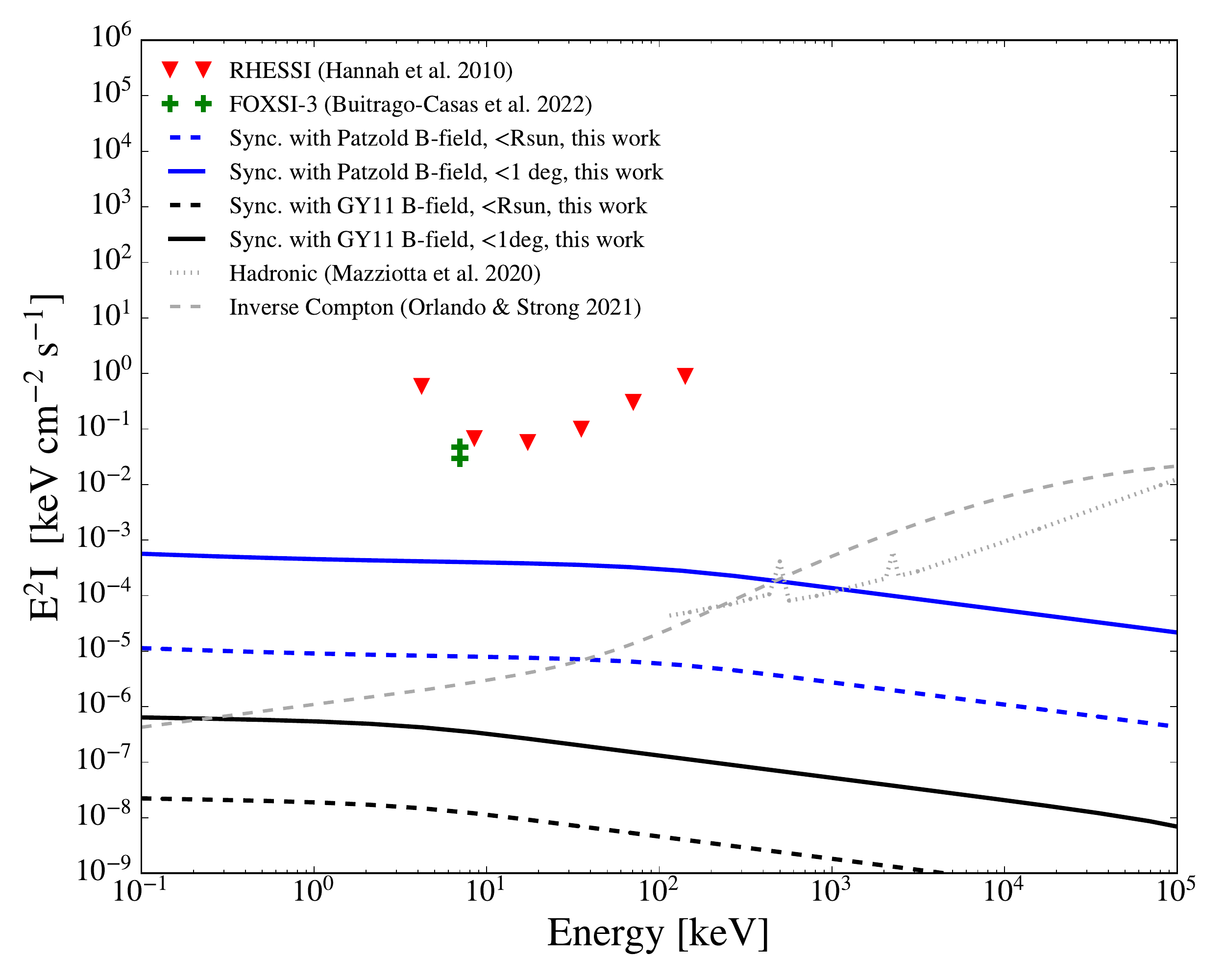}
\caption{Calculated synchrotron spectral flux with the Patzold (blue lines) and GY11 (black lines) models. Dashed lines are obtained by integrating the emission in the entire solar disk; solid lines are obtained by integrating the emission in a disk of 1 degree from the center of the Sun. Red triangles are RHESSI flux upper limits for the entire solar disk. Green crosses are the FOXSI-3 upper limits of the quiet Sun. Also shown are the inverse-Compton \citep{Stellarics} (gray dashed line) and the pion decay \citep{Mazziotta} (gray dotted line) model components.
\label{figure3}}
\end{figure}
To better investigate this effect and to understand which solar regions contribute the most to the entire flux, we show the calculated synchrotron intensity profile along the line-of-sight as a function of the angular distance from the center of the Sun (see Figure~\ref{figure4}) for the two magnetic field models. 
\begin{figure} 
\centering
\includegraphics[width=0.6\textwidth, angle=0]{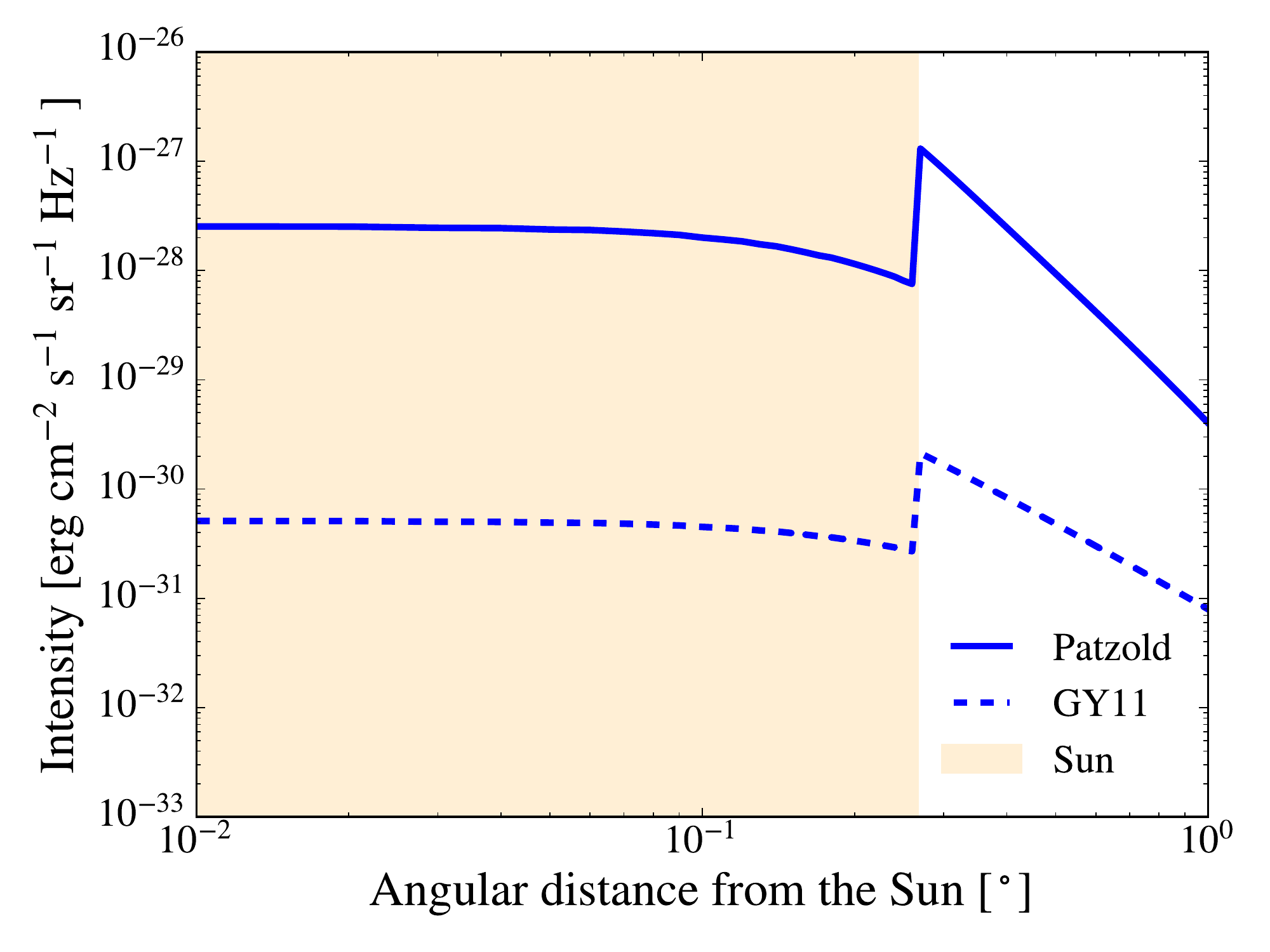}
\caption{X-ray intensity profile of the synchrotron emission calculated at 10$^{18}$ Hz with Patzold model (solid line) and with GY11 model (dashed line) of the magnetic field reported in Figure~\ref{figure2}. The yellow region identifies the solar disk extension.
\label{figure4}}
\end{figure}
We find that, while the synchrotron intensity is almost constant in the solar disk, it significantly increases at the solar limb, and, then, it quickly drops away from the Sun. 
This intensity profile is explained by a combination of two factors: the fact that CR electrons do not penetrate inside the Sun, and the fact that the magnetic field decreases with the distance from the Sun. The quick drop is explained by the following.
Because the intensity of the magnetic field drops  with the distance from the Sun, the synchrotron intensity
quadratically depends on the magnetic field intensity,  as seen above. Hence, the radiation is more concentrated in the close proximity of the Sun. 
The results are that the magnetic field very close to the Sun mostly contributes to the synchrotron flux, and that the synchrotron emission quickly drops away from the Sun.
We find the integrated flux of synchrotron emission within the solar disk to be about 25\% and 15$\%$ of the total flux from the Sun for the Patzold and GY11 models respectively, while the region outside R$_{sun}$ makes almost 75$\%$ and 85$\%$ of the total flux respectively.

\section{Discussion}
\subsection{RHESSI heredity and predictions for NuSTAR and FOXSI}
RHESSI \citep{RHESSI2010} has placed the lowest upper limits on the hard
X-ray emission from the quiet Sun from 3 to 200 keV, allowing strong
constraints on any energy released by nanoflares  that fall below the RHESSI detection limit and even on axions possibly produced in the solar core. Improvements on the observed upper limits of the quiet Sun depend also by the sensitivity of the instrument to detect small flares.
RHESSI has detected more than 25,000 transient X-ray microflares in the 6–12 keV energy band. Due to their spatial distribution from active regions, it is established that they do not heat the quiet corona \citep{Christe, Hannah2008}. NuSTAR observed the Sun several times since the start of solar pointing in September 2014. Authors \citep{Marsh} state that NuSTAR is sensitive to transient events several times smaller than the RHESSI detection threshold for identical integration times. During solar minimum they expects the NuSTAR sensitivity to increase by over two orders of magnitude due to higher instrument livetime and reduced solar background. 
Its minimum detectable flux for 1Ms of 1.7 $\times$ 10$^{-6}$, 2.9 $\times$ 10$^{-6}$, 1.0 $\times$ 10$^{-4}$, 8.3 $\times$ 10$^{-3}$ keV cm$^{-2}$ s$^{-1}$ at 3, 10, 30, and 78 keV respectively (F. Harrison and K. Madsen, private communication) is comparable to synchrotron intensities shown in Figure 3.
However, even though the NuSTAR instrument has higher sensitivity than RHESSI, it also handles a limited amount of photons at a time (I. Hannah private communications). As a result, many photons may come from the corona, hence, limiting full-disk observations. 
On the other side, a FOXSI-type solar mission can both produce direct imaging like NuSTAR and also it can handle the larger solar flux. FOXSI is the first solar-dedicated instrument to observe hard X-rays with focusing optics. FOXSI-4\footnote{https://foxsi.umn.edu/launches/foxsi-4}, scheduled for 2024, will continue to improve upon direct solar imaging.
A possible future FOXSI’s concept would allow enough observation time to further constrain the current hard X-ray quiet Sun limits.
The quiet Sun may be observed in various solar activity conditions and the synchrotron emission could be detected and studied in detail. 
The observed solar profile  might be obtained and compared with our model expectations of the synchrotron emission. 
By observing the synchrotron solar profile CR density and magnetic field intensity models  could be  tested and their intensity at the Sun could be indirectly traced.
Additionally, the synchrotron emission from GCR needs also to be accounted for when searching for axions \citep{Sikivie}, which are hypothesized dark matter particles from the Sun. 
Solar axions are hypothesized to be produced in the solar core via nuclear reactions and to be converted to X-rays in a presence of a strong magnetic field \cite[e.g.][]{HudsonAxions}. RHESSI has been used to search for faint X-ray emission from axions converting in coronal magnetic fields \cite[e.g.][]{HannahAxions}.

\subsection{Synchrotron modeling and observations from Radio to UV}
In this section we calculate the synchrotron intensity by GCR from radio to UV and we compare it with observations.
Recently, the Low-Frequency Array (LOFAR) \citep{Lofar} measured the solar brightness temperature ranging from $\sim$10$^5$ to $\sim$10$^6$~K in the 20–80 MHz range. 
As previously found, the size of the Sun measured by LOFAR 
increases with the decrease of the frequency, with a radius larger than the local plasma frequency radius.
In fact, it is known that at very low frequencies (below $\sim$0.1 GHz) the solar disk appears much bigger than its optical size, and its brightness gradually decreases and vanishes after several solar radii. 
For frequencies above several GHz the size of the Sun appears similar to its visible counterpart. Hence, higher frequency radio emission originates closer to the photosphere, while lower frequency radio emission originates in the corona, which gives the Sun a larger dimension in the sky.
In the following, we start by reporting and discussing our estimates of the synchrotron emission in radio and, by comparing it with current LOFAR observations, we test if synchrotron emission in the radio band can contribute to the larger radius seen at low frequencies. 
The GCR electrons  producing synchrotron emission from a few MHz to 100 MHz have energies  up to 100 MeV. 
Hence, we calculate the synchrotron emission by accounting for CR electron measurements at low energies from Balloons \citep{Balloons1, Balloons2} and from PAMELA \citep{Pamela}. 
The resulting synchrotron brightness temperature has an almost constant temperature in the disk, a higher temperature in the limb, and it drops with the angular distance from the Sun with a similar shape as found in X-rays. We find it to be 5 - 6 orders of magnitude lower with respect to the latest observations by LOFAR \citep{Lofar},  hence, not significant in the radio domain, and we avoid to plot it.
Because the expected brightness temperature of the synchrotron emission is decreasing with the increase of the frequency, this emission from GCR cannot be  distinguished in the radio band,  neither with current very low frequency telescopes such as the forthcoming space mission The Sun Radio Interferometer Space Experiment (SunRISE) \citep{SUNRISE}.
Moreover,  while at 1 MHz the expected synchrotron brightness temperature in the proximity of the Sun reaches a few hundred K for Patzold model, at 10 MHz is below or at the same level of the cosmic microwave background. For comparison the brightness temperature of the Galactic synchrotron background is $\sim$10$^6$~K at 1 MHz \citep{OS2013}. As a result, in the radio band the synchrotron emission from the Sun can be considered negligible with respect to the other emission mechanisms. 
Recently, the Atacama Large Millimeter and submillimeter Array (ALMA), working in the frequency band 35~-~950~GHz, was also able to image the quiet Sun \citep{ALMA}. However, it measured a brightness temperature around 6000~K, which is more than 11 orders of magnitude higher than what we find from the synchrotron emission by GCR. At even lower wavelengths than ALMA, up to UV, the calculated expected synchrotron emission from the Sun is totally insignificant compared to the brightness temperature of the solar thermal radiation.

\subsection{GCR heliopheric propagation and energy losses}
Our current calculations of the synchrotron spectrum by GCR can be considered as an upper limit to this emission, given the fact that we did not account for GCR energy losses.
Here, we discuss the expected effect on our estimates if energy losses are considered. Energetic GCR may lose energy by synchrotron and inverse Compton emission. A reduction of the intensity of the GCR electrons at the Sun would linearly reflect on a reduction of the intensity of the expected synchrotron emission.
Uncertainties on the  energy losses are very large and strongly depend on the choice of the transport parameters. Calculations of electron propagation in the heliosphere including diffusion and energy losses is beyond the present effort and reported elsewhere \citep{POS}. 
Moreover, in additional further works we aim at eventually constraining energy losses parameters with improved detection of the quiet Sun in gamma rays originated by a different emission mechanism, such as the significantly observed inverse Compton emission \citep[e.g.][]{Stellarics}.
Based on these considerations and based on magnetic field models that are built on present observations, we expect the solar synchrotron intensity from radio to gamma rays being equal or lower than estimated here. Note that this does not change our conclusions.

\section{Summary}
Starting from current precise GCR electron measurements and recent solar and Heliospheric magnetic field observations, we have reported estimates of the synchrotron emission from gamma rays to radio in the direction of the Sun and in the Heliosphere produced by GCR electrons in the solar magnetic field. To the best of our knowledge this is the first time this emission component has been proposed and modeled. 
We have analyzed the expected synchrotron profile 
and found that it is almost constant within the solar disk, while it peaks in the very proximity of the Sun where the magnetic field is maximum, then to rapidly fall away from the solar surface. 
By comparing our estimates with recent upper limits of the quiet solar emission by RHESSI in X-rays we have found that the expected emission component is a few orders of magnitude lower than current upper limits. 
We have concluded that, even though it does not explain current RHESSI upper limits, this new emission component provides a more complete description of the quiet Sun and   could be observed and studied in the near future. This   could be promising for NuSTAR and, especially, for future FOXSI observations. We have also compared our calculation of the synchrotron brightness temperature from the Sun with latest observations of the quiet Sun in radio, especially with LOFAR, finding that the upper limits of the expected synchrotron emission component is of a few K, i.e. 5~-~6 orders of magnitude lower than the brightness temperature currently measured in the 10~-~100~MHz range from the quiet Sun. 
We have also found that, from millimeters to UV,  the synchrotron emission component is  negligible compared to the brightness temperature of the solar thermal radiation. Hence, because the expected synchrotron emission is several orders of magnitude lower than what is observed from radio to UV, we have concluded that there is little prospect for its detection in this band. On the other side, we have showed that this emission could, in future, potentially be observed at high energies.

\begin{acknowledgments}
The authors thank I. Hannah, F. Harrison, J. Lazio, and K. Madsen for useful comments. E.O. acknowledges the ASI-INAF agreement n. 2017-14-H.0, the NASA Grant No. 80NSSC20K1558. V.P. is supported by NASA Living With Star program grant NNH20ZDA001N-LWS
\end{acknowledgments}

\end{document}